\documentclass[aps,twocolumn,amsmath,amssymb,prA]{revtex4}
\usepackage{bbm}
\usepackage{amsfonts} 
\usepackage{amssymb}
\usepackage{amsmath}
\usepackage[dvips]{graphicx}
\usepackage{epsfig}
\newcommand{\beq}{\begin{equation}}
\newcommand{\eeq}{\end{equation}}
\newcommand{\bee}{\begin{eqnarray}}
\newcommand{\eee}{\end{eqnarray}}
\newcommand{\been}{\begin{eqnarray*}}
\newcommand{\eeen}{\end{eqnarray*}}
\newcommand{\da}{\dagger}

\newcommand{\bmm}{\begin{matrix}}
\newcommand{\emm}{\end{matrix}}

\newcommand{\up}{\uparrow}
\newcommand{\dn}{\downarrow}
\newcommand{\tik}{\tilde{k}}

\begin{document}

\input epsf.sty

\title{Inversion symmetry breaking and criticality in free fermionic lattices}

\author{Zolt\'an K\'ad\'ar}\email{zokadar@gmail.com}
\affiliation{School of Mathematics, University of Leeds, Leeds, LS2 9JT, United Kingdom}
\date{\today}

\begin{abstract}
%
We describe the connection between inversion symmetry breaking and criticality in free fermionic lattice models. It is shown that for translation-invariant spinless fermions, the breaking of this symmetry in the ground state implies criticality, i.e., the existence of long-range correlations and the vanishing of the spectral gap; while for models with spin, only the asymmetry of the spin-averaged covariance  matrix  implies a similar conclusion. Our results are proved by introducing invariants under global translation-invariant free fermion quenches. Using this result, we identify a set of models where the generalized Hartree-Fock approximation must break down.
\end{abstract}

\maketitle

\section{Introduction}

Symmetries play a prominent role in the characterization of many-body systems. A very timely topic in this respect is, for example, the theory of topological insulators and superconductors \cite{Qirev}. These phases of matter are modelled on free fermion lattices, gapped and 
possess a topological invariant that does not change its value if one adiabatically changes the system by local perturbations that preserve certain symmetries and do not close the gap. In this paper, we describe a dual result: we introduce quantities that are invariant under sudden global (translation-invariant) quenches, and whose non-zero expectation values imply criticality. 


The introduced invariants are connected to the inversion symmetry breaking of the ground state. In our set-up the free fermion Hamiltonians explicitly break this symmetry, and we show that they can only be gapped if the expectation values of these invariants vanishes in the ground state. 

Inversion symmetry breaking lattice models appear in the realm of many physical scenarios, e.g., in non-equilibrium states 
\cite{EZ1,EZ2}, in the case of Dzyaloshinskii-Moriya interactions \cite{DerMo,DVKB,JKLS,LL}, or in directed quantum transport \cite{TRSBQT}. In a previous work, we noticed that when the covariance matrix of
a translation-invariant spinless free-fermion chain breaks reflection symmetry (which is identical to inversion in one dimension), 
it is necessarily 
gapless \cite{mi}. This finding lead us to study whether a similar statement holds
with spin degrees of freedom and/or in higher-dimensions. 
Using the structure of the general Hamiltonian describing  translation-invariant  $d$ dimensional spinful 
quasifree models, we introduce a set of invariants connected to inversion symmetry breaking
that can be used to signal criticality. Concerning topological insulators and superconductors, this result implies that the
models are either interacting or inversion symmetry breaking happens at the quantum critical point \cite{tai,wchf,kf,sri}.

When the expectation values of the mentioned invariants do not vanish in a 
translation-invariant quasifree state, 
the area law for the entanglement entropy is
logarithmically violated.
An immediate consequence of this is the breakdown of
the generalized Hartree-Fock method and band model approximations
for certain  interacting gapped models with inversion symmetry breaking.

%
%
%
%
%

%
%
%
%
%
%
%

The paper is structured as follows. In Section II we introduce quasifree fermion models,
and provide their  ground-state two-point function in full generality. The main result is stated
and proved in Section III, after an example of a gapped pairing model, which
shows that the naive generalization of the spinless invariants does not work.
Finally, the insufficiencies of generalized Hartree-Fock approximation is explained.

%
%
%
%

\section{Reflection symmetry in free fermion Hamiltonians}

\subsection{The model}

We consider a $d$ dimensional cubic lattice of fermions with arbitrary spin.
The fermion operator $b^j_n$ is the annihilation operator at the lattice point denoted by 
$n\equiv (n_1,n_2,\dots,n_d)$, where $n_i\in\{1,2,\dots,N_i\}$, and the spin index $j\in \{1,2,\dots,s\}$,
the creation operators $b^{j\dagger}_n$ have analogous notation. The most general quadratic combination of 
these fermion operators yielding a Hermitian operator on the Fock space reads

\beq H
{=} \sum_{j,l=1}^{s}\sum_{m,n=1}^NA^{jl}_{mn} b_m^{j\dagger}b_n^l 
{+}\frac{1}{2}\left(
B_{mn}^{jl} b_{m}^{j\dagger}b_{n}^{l\dagger}-
\overline{B}_{mn}^{jl}b_m^j b_{n}^l\right)
 , \label{Hqfree}
\eeq
where summation of $m$ and $n$ is over the sites $N$ of the lattice (e.g., for a square lattice in dimension two, the multi-index $m\equiv(m_1,m_2)$ labels all $N\equiv N_1 N_2$ sites in the natural way). We assume periodic boundary conditions. Without loss of generality one can impose that
$B^{jl}_{mn}= -B^{jl}_{nm}$, and the constraint $H^\dagger=H$ is equivalent to $A^{jl}_{mn}=\overline{A}^{lj}_{nm}$.
%

We shall often write the $Ns$ dimensional
vector $b$ ($b^\da$) without indices, whose components are 
$b_m^j$ ($b_m^{j\da}$, respectively). 
It is customary to write the above Hamiltonian as
%
\beq H=\begin{array}{ll}\frac{1}{2}(b^\dagger&b)\\\quad\end{array}
\left(\begin{array}{rc}A&B\\-\overline{B}&-A^T\end{array}\right)
\left(\begin{array}{l}b\\b^\dagger\end{array}\right) ,\label{BdG}\eeq
where the big matrix of size $2sN \times 2sN$ is
called the Bogoliubov--{de~Gennes} (BdG) Hamiltonian.
 
Translation invariance implies that all the coefficient matrices are circulant:
$X^{jl}_{m+p,n+p}=X^{jl}_{m,n}$ for any translation $p\equiv(p_1,p_2,\dots,p_d)$ (addition, 
subtraction and scalar multiplication between multi-indices are naturally meant componentwise). 
The periodic boundary condition is taken into account through the definitions $b^j_n=b^j_{n+\underline{N}}$, with $\underline{N}\equiv (N_1,N_2,\dots,N_d)$.

\subsection{Two-point function of the ground state}
Let us fix our conventions used in the calculation. 
The Fourier of the one-particle annihilation operators  and its inverse 
read (we use the variable $k$ or $k'$ exclusively 
for the Fourier transform in what follows)
\bee b^j_k&=&\frac{1}{\sqrt{N}}
\sum_n\exp\left(-\frac{2\pi i n k}{N}\right) 
b^j_n,\label{fourm}\\
b^j_n&=&\frac{1}{\sqrt{N}}\sum_k\exp\left(\frac{2\pi i n k}{N}\right) 
b^j_k,
\eee
where $N\equiv \prod_i^d N_i$, $k\equiv(k_1,k_2,\dots k_d)$, $nk\equiv\sum_i^d n_i k_i$ 
and there are $d$ summations over $n_i$ or $k_i$, $i=1,2,\dots d$, that is, 
the $i$-th component of $n$ and $k$ run in from $1$ to $N_i$. The transform of the one-particle
creation operators are to be computed by means 
of taking the adjoint of the above formulae.

For circulant matrices we define the Fourier transform as
\bee
     X^\xi_k&=&\sum_n\exp\left(-\frac{2\pi i n k}{N}\right)X^\xi_{n\bf{0}} \, ,\\
     X^\xi_{n\bf{0}}&=&\frac{1}{N}\sum_k\exp\left(\frac{2\pi i n k}{N}\right)X^\xi_k \, ,
\eee
where $\xi$ stands for the pair of spin indices
($\bf{0}$ is the $d$ dimensional zero vector).  Summation conventions are identical to the above.

Using these definitions, the Hamiltonian (\ref{Hqfree}) or (\ref{BdG}) can be written as
\beq H=\frac{1}{2}\sum_k\begin{array}{ll}(b_k^\dagger&b_{-k})\\\quad\end{array}
\left(\begin{array}{rc}A_k&B_k\\B^\dagger_k&-A_{-k}^T\end{array}\right)
\left(\begin{array}{l}b_k\\b_{-k}^\dagger\end{array}\right)\ ,\label{BdGk}\eeq
where the $2s \times 2s$ big matrix is denoted by ${\cal H}_k$, and is called 
the Bogoliubov-{de~Gennes} (BdG) Hamiltonian.
The hermiticity constraint
imply 
\beq
A_k^\dagger=A_k, \quad B_{-k}=-B_k^T,
\eeq
and in particular, $A_k^{jj}\in {\mathbb R}$ and $B_{-k}^{jj}=-B_k^{jj}$.
%

To bring this Hamiltonian into a diagonal form 
\[ H=\sum_{j=1}^s\sum_k \Lambda^j_k\,c^{j\da}_k c^j_k,\;\;(\Lambda^j_k\in{\mathbb R})\ ,\] 
one performs a Bogoliubov transformation \footnote{This is the most general translation-invariant
Bogoliubov transformation.} 
\beq c^j_k=\sum_{l=1}^{s}\left(\alpha^{jl}_k b^l_k+\beta^{jl}_k b^{l\dagger}_{-k}\right),\quad\alpha^{jl}_k,
\beta^{jl}_k\in {\mathbb C}\ ,\label{bt}\eeq 
where the coefficients $\alpha^{jl}_k,\beta^{jl}_k$ have to satisfy for each $j,j'\in\{1,2,\dots,s\}$
\bee \sum_{l=1}^s\left(\alpha^{jl}_k\beta^{j'l}_{-k}+\beta^{jl}_k\alpha^{j'l}_{-k}\right)&=&0,\label{bogocona}\\
\sum_{l=1}^s\left(\alpha^{jl}_k \overline{\alpha}^{j'l}_k+\beta^{jl}_k\overline{\beta}^{j'l}_k\right)
&=&\delta_{jj'}\ ,\label{bogoconb}\eee    
so that the canonical anticommutation relations 
$\{c^j_k,c^{j'\dagger}_{k'}\}=\delta_{jj'}\delta_{kk'},\;\{c^j_k,c^{j'}_{k'}\}=0$ are satisfied. 
The consistency conditions for the commutator 
\[[c^j_k(b),H(b)]=\Lambda^j_k\,c^j_k(b)\]
 yield the eigenvalue 
equations\footnote{The calculation of the consistency condition gives the eigenvalue equation for the block matrix 
$\left({A_k^T,\;\;\;\overline{B}_k \atop B_k^T,-A_{-k}}\right)$ with eigenvectors $\overline{v}_k^j$. The equations are
equivalent to the eigenvalue equations of ${\cal H}_k$ written in the main text.}
\beq {\cal H}_k v^j_k=\Lambda^j_k v^j_k \label{eequ}\eeq
 with $v^j_k\equiv
(\overline{\alpha}^{j1}_k,\overline{\alpha}^{j\,2}_k,\ldots,\overline{\alpha}^{j\,s}_k,
\overline{\beta}^{j\,1}_k,\overline{\beta}^{j\,2}_k,\ldots,\overline{\beta}^{j\,s}_k).$
Let us use the notations $\Psi_k\equiv (b_k,b_{-k}^\da)$ and 
$\tilde{\Psi}_k\equiv (c_k,c_{-k}^\da)$ to write $2H=\Psi_k^\da {\cal H}_k \Psi_k=\tilde{\Psi}_k^\da {
\cal H}^d_k \tilde{\Psi}_k$ with ${\cal H}^d_k$ diagonal matrix. 
The generic BdG Hamiltonian satisfies 
\beq \sigma^{ph}_x {\cal H}_k\sigma^{ph}_x=-\overline{\cal H}_{-k}\label{phs}\eeq 
with
$\sigma^{ph}_x$ being the first Pauli matrix acting in the ``particle-hole" space (that is, the indicated splitting
of the $2s$ by $2s$ matrix into $s$ by $s$ blocks). The property~\eqref{phs} is sometimes called particle-hole symmetry. Note, 
that this is {\em always} present in translational invariant quasifree fermion systems. As a consequence,
the form of the diagonal Hamiltonian reads   
${\cal H}^d_k=\mbox{diag}(\Lambda_k^1,\Lambda_k^2,\dots,\Lambda_k^s,-\Lambda_{-k}^1,
-\Lambda_{-k}^2,\dots,-\Lambda_{-k}^s)$. The unitary defined by $\Psi_k=U\tilde{\Psi}_k$ can be read off from the inverse
of (\ref{bt})
\beq b^j_k=\sum_{l=1}^s\left(\overline{\alpha}^{lj}_k c^l_k+\beta^{lj}_{-k}c^{l\da}_{-k}\right),\label{inve}\eeq
it reads
\beq U=\left(\begin{array}{rr}\alpha_k^\da&\beta_{-k}^T\\\beta_k^\da&\alpha_{-k}^T\end{array}\right)\label{U}\eeq
and the eigenvectors of ${\cal H}_k$ are its columns by definitions.

In the ground state, the two-point functions of the new 
Fermi operators read 
\bee \langle c_k^{j\dagger}\,c^{j'}_{k'} \rangle &=& \frac{1}{2}
\left(-\frac{\Lambda^j_k}{|\Lambda^j_k|} +1 \right)\delta_{j,j'}\delta_{k,k'},\\
\langle c_k^{j}\,c^{j' }_{k'} \rangle  &=&\langle c_k^{j\dagger}\,c^{j' \dagger}_{k'} \rangle =0.
\eee
Using these we can express those of the original ones using (\ref{inve}) (summation over the Fourier $k$ and the spin spin index $l$) 
\begin{eqnarray}
\langle b^j_m\,b^{j'}_{\,n} \rangle\!\!&&=\!\!\frac{1}{2N}
\sum_{k,l}\exp\frac{2\pi ik(m-n)}{N}\,(bb)^{l,j,j'}_k\nonumber
 \\
\langle b^{j\da}_m\,b^{j'}_{\,n} \rangle\!\!&&=\!\!\frac{1}{2N}
\sum_{k,l}\exp\frac{2\pi ik(m-n)}{N}\,(b^\dagger b)^{l,j,j'}_k \nonumber
\end{eqnarray}
with the kernels 
\begin{eqnarray*}
(bb)^{l,j,j'}_k&\equiv&\overline{\alpha}^{lj}_k\beta^{lj'}_k
\left(\frac{\Lambda^l_k}{|\Lambda^l_k|} +1 \right)
+\beta^{lj}_{-k}\overline{\alpha}^{lj'}_{-k}
\left(-\frac{\Lambda^l_{-k}}{|\Lambda^l_{-k}|} +1 \right)\\
(b^\dagger b)^{l,j,j'}_k&\equiv&
\alpha^{lj}_{-k}\overline{\alpha}^{lj'}_{-k}
\left(-\frac{\Lambda^l_{-k}}{|\Lambda^l_{-k}|} +1 \right)
+\overline{\beta}^{lj}_k \beta^{lj'}_k
\left(\frac{\Lambda^l_k}{|\Lambda^l_k|} +1 \right)
\end{eqnarray*}

Now, let us introduce the following notations
\[
M^l_k\!=\!\frac{1}{2}\!\left(\frac{\Lambda^l_k}{|\Lambda^l_k|}
-\frac{\Lambda^l_{-k}}{|\Lambda^l_{-k}|}\right),\quad
P^l_k\!=\!\frac{1}{2}\!\left(\frac{\Lambda^l_k}{|\Lambda^l_k|}
+\frac{\Lambda^l_{-k}}{|\Lambda^l_{-k}|}\right),\]
\[ (S^{l\pm}_k)_{jj'}\!=\!\overline{\alpha}_k^{lj}\beta_k^{lj'}
\!\pm\beta_{-k}^{lj}\overline{\alpha}^{lj'}_{-k},\; (Z^{l\pm}_k)_{jj'}\!=\!\overline{\alpha}_k^{lj}\alpha_k^{lj'}
\!\pm\beta_{-k}^{lj}\overline{\beta}^{lj'}_{-k}\ ,\]
in terms of which the kernels defined above can be conveniently written as
\begin{eqnarray}
(bb)^{l,j,j'}_k=(M_k^l+1)(S^{l+}_k)_{jj'}+P_k^l (S^{l-}_k)_{jj'}\label{ccd1}\\
(b^\dagger b)^{l,j,j'}_k=(M_k^l+1)\overline{(Z^{l+}_{-k})}_{jj'}-P_k^l (\overline{Z^{l-}_{-k})}_{jj'}
\label{ccd2} \end{eqnarray}
Before analysing these results from the point of view of inversion symmetry breaking, we write
down the following useful set of identities:
\begin{eqnarray} \sum_{l=1}^s S_k^{l+}&=&{\bf 0}, \nonumber
\\\sum_{l=1}^s (Z_{k}^{l+})_{jj'}&=&\delta_{jj'}\ .\label{ort2} \end{eqnarray} 
They are the components of the matrix equation $UU^\da={\mathbbm 1}$ with $U$ being the 
unitary defined by (\ref{U}) 
(c.f., the components of the equation $U^\da U={\mathbbm 1}$ are equivalent to \eqref{bogocona} and \eqref{bogoconb}).

\subsection{(Extended) inversion symmetry}
The inversion symmetry transformation whose breaking is related to criticality is given by the transformation 
\[b_m^j\mapsto i b_{-m}^j
\] 
implying $A^{jl}_{mn}\mapsto \overline{A}^{lj}_{mn}$ and no change in the pairing coefficients $B_{mn}^{jl}$. Or 
using the notation $A_{mn}$ for the $s \times s$ matrix, whose entries are given by $(A_{mn})_{jl}=A_{mn}^{jl}$ we can express
the transformation as $A_{mn}\mapsto A^\da_{mn}$.

In case the model is spinless the formula from \cite{mi}
relating the one-particle spectrum to the coeffients of the BdG Hamiltonian%
\beq \Lambda_k=\frac{A_{k}-A_{-k}+
\sqrt{(A_k+A_{-k})^2+4B_k\overline{B}_k}}{2} \, ,
\label{ops}\eeq
applies (with k standing for the coordinate of the $d$ dimensional momentum torus). 
In other words, since $A_k-A_{-k}=\Lambda_k-\Lambda_{-k}$ and $A_k-A_{-k}\neq 0$ means broken inversion symmetry, all we have 
to investigate is the dependence on the k-antisymmetric part of the one-particle speactrum. 

In the mentioned previous work \cite{mi}, the starting point was a quasifree spin chain, which by definition can be 
transformed by Jordan-Wigner transformation to a fermion chain. There, studying inversion symmetry breaking 
we arrived at the above conclusion and investigated when the ground state is 
sensitive to $(\Lambda_k - \Lambda_{-k})\neq 0$ \footnote{Let us note that due to the 
non-local nature of the Jordan-Wigner transformation, the reflection symmetry transformation 
in the spin chain is not identical to that in the fermion model.}.
We have found that only the imaginary part of the two-point 
functions $\langle b_j^\da b_l \rangle$ depend on this quantity and
that dependence appears via the combination $M_k$ 
$\sim(\Lambda_k/|\Lambda_k|-\Lambda_{-k}/|\Lambda_{-k}|)$. If there is a momentum $k_0$ with $M_{k_0}\neq 0$, 
this implies that $\Lambda_{k_1}=0$ at some 
momentum $k_1$. Consequently, the gap disappears whenever $Im\langle b_j^\da b_l \rangle\neq 0$
 (and the corresponding spin-chain ground state 
breaks inversion symmetry).

For the general quasifree fermion lattice we cannot express $\Lambda_k^j$ explicitly in terms of the coefficients of the Hamiltonian 
(the solution of the characteristic equation of the $2s \times 2s$ BdG matrix). But we can investigate if 
 the non-vanishing of the quantity $Im\langle b_m^{\da j} b_n^l\rangle$ or a suitably modified version of it signals inversion symmetry
breaking and if it leads to criticality.

\section{The main result}

In higher dimensional and spinful quasifree fermionic lattices the 
non-vanishing of $Im\langle {b^j_m}^\da b^{j}_n \rangle$\footnote{The correlator between different spin components is easily excluded: 
suitable simple Bogoliubov transformation results in non-real expectation value in a gapped ground state of spinful quasifree model.} 
does {\em not} imply criticality: it is clearly demonstrated
by the following simple model. Consider consider the following nearest-neighbor 
spin-$\tfrac{1}{2}$ Hamiltonian given by
%
%
\[\begin{array}{ll}&{\displaystyle H=\frac{1}{2}\sum_m}{\displaystyle \Big[(p-1)(b_m^{\up\da}b_m^\up+b_m^{\dn\da}b_m^\dn)}+\\\\
&{\displaystyle i\,\frac{p+1}{2}(-b_m^{\up\da}b_{m+1}^\up+b_{m+1}^{\up\da}b_m^{\up}+b_m^{\dn\da}b_{m+1}^\dn-b_{m+1}^{\dn\da}b_m^{\dn})}\\\\&
{\displaystyle
-\frac{p+1}{2}(b_m^{\up\da}b_{m+1}^\dn+b_{m+1}^{\up\da}b_{m}^\dn+b_m^{\dn\da}b_{m+1}^\up+b_{m+1}^{\dn\da}b_{m}^\up)\Big]}

\end{array}\]
We assume that the parameter $p>0$.
In the momentum space (with the abbreviation $\tilde{k}=2\pi k/N$) it has the form
\[\begin{array}{l} {\displaystyle H\!=\!\sum_k \frac{p+1}{2}\sin \tilde{k}\big(b_k^{\up\da}b_k^{\up}-
b_k^{\dn\da}b_k^{\dn}\big)+\!\frac{p-1}{2}\!\big(b_k^{\up\da}b_k^{\up}+b_k^{\dn\da}b_k^{\dn}\big)}\\\\{\displaystyle
-\frac{p+1}{2}\cos \tik\, 
(b_k^{\up\da}b_k^\dn+
b_k^{\dn\da}b_k^\up)}\ .\end{array}\]
The diagonal form $H=\sum_k(-c^{\up\da}_kc^\up_k+p\,c^{\dn\da}_k c^{\dn}_k)$ is obtained by the transformation
\begin{eqnarray*} b_k^\up&=& \frac{1}{\sqrt{2}}\left((\cos\tik/2-\sin\tik/2)\,c_k^{\up}+
(\cos\tik/2+\sin\tik/2)\,c_k^{\dn}\right)\\
b_k^\dn&=&\frac{1}{\sqrt{2}}\left((\cos\tik/2+\sin\tik/2)\,c_{k}^{\up}-(\cos\tik/2-\sin\tik/2)\,c_k^\dn\right),\end{eqnarray*}
%
Now we can compute arbitrary two-point functions.
In particular, we get 
\[\langle b_m^{\up\da} b_{m+1}^\up\rangle=-\frac{i}{4}\quad\langle b_m^{\dn\da} b_{m+1}^\dn\rangle=\frac{i}{4}\ ,\] 
showing that these quantities need not be real,  although the model is gapped.

However, we found the proper generalization of the spinless one-dimensional result:\\

\noindent {\bf Main result.} {\it Consider the ground state of a model given by the Hamiltonian~\eqref{Hqfree}, and let $U$ be a unitary implementing an arbitrary translation-invariant 
Bogoliubov transformation. Then the following hold: (i)
\begin{equation}Im( \sum_{j=1}^s \langle {b^j_m}^\da b^{j}_{n} \rangle)= 
 Im( \sum_{j=1}^s \langle U{b^j_m}^\da b^{j}_{n}U^{\dagger} \rangle)\ ,\label{rinv}\end{equation}
i.e., the lhs. of (\ref{rinv}) is $U$-invariant;\\
(ii) if  
\[Im( \sum_{j=1}^s \langle {b^j_m}^\da b^{j}_{n} \rangle)\neq 0\] for certain  
$m$ and $n$, then the model is gapless.}\\

%
%
%

We will prove point (ii) of the proposition in the next subsection using directly the parameters of  the diagonalization in Sec.~II. Later in Sec.~III.B we present a more abstract treatment, sketching a proof of (i) and through that an alternative simple proof of (ii).

%
%
%

\subsection{Direct parametrization method}

To prove the proposition, we notice from the form of Eq.~\eqref{ccd2} that if $M_k^l\equiv 0$ 
for all spin components $l$, then, without loss of generality we can assume that $P_k^l\equiv 1$.\footnote{
The signs of $\Lambda_k^j$ and $\Lambda_{-k}^j$ may be simultaneously negative, but the translation-invariant Bogoliubov 
transformation $c^j_k\leftrightarrow c^{j\da}_{-k}$ on the diagonal BdG Hamiltonian changes the sign of $P_k^j$.}
We first write -- using the identities (\ref{ort2}) -- 
the Fourier kernel of the two-point function (\ref{ccd2}) as 
\begin{align*}\sum_{l=1}^s (b^\dagger b)^{l,j,j}_k&=1\!-\!\sum_{l=1}^s \overline{(Z^{l-}_{-k})}_{jj}=1\!+\!\sum_{l=1}^s 
(-|\alpha^{lj}_{-k}|^2\!+\!|\beta^{lj}_k|^2)\\
&=1
-\sum_{l=1}^s (|\alpha^{lj}_{-k}|^2+|\beta^{lj}_k|^2)+2\sum_l|\beta^{lj}_k|)\\&=2\sum_{l=1}^s |\beta_k^{lj}|^2\ ,
\end{align*}
where in the last equality we used (\ref{ort2}) again. Now we need the summed version of two equation 
(\ref{bogoconb}) and (\ref{ort2}) to arrive at
\[ \sum_{l,j=1}^s(|\alpha_k^{jl}|^2+|\beta_k^{jl}|^2)=s=\sum_{l,j=1}^s(|\alpha_k^{jl}|^2+|\beta_{-k}^{jl}|^2)\ .\]
This shows that $\sum_{l,j}|\beta_k^{lj}|=\sum_{l,j}|\beta_{-k}^{lj}|$, which implies that the summed Fourier kernel
$\sum_{l,j}(b^\dagger b)^{l,j,j}_k$ is also symmetric in $k$. But it is also real, so the quantity 
$\sum_j \langle b_m^{j\da}b_n^j\rangle$ is 
also real. 

This means that if $Im(\sum_j \langle b_m^{j\da}b_n^j\rangle) \ne 0$, then at least 
for one $j$ and $k_0$ we have $M_{k_0}^j = \tfrac{1}{2} (\Lambda^j_{k_0}/|\Lambda^j_{k_0}|-\Lambda^j_{-k_0}/|\Lambda^j_{-k_0}|)\ne 0$,
which, in turn, implies
the absence of the gap as explained in the beginning of the section for 
the spinless case \footnote{Note gaplessness follows here for the decoupled band $j$, 
and consequently for the whole model.}.
\subsection{Method of commuting invariants}
We start by intruducing a general Bogoliubov transformation, which is a basis change in the space
of creation and annihilation operators
\[ b_m^j\mapsto {\cal U} b_m^j {\cal U}^\da\]
which keep the canonical anticommutation relations invariant. Now we can turn to the proof, 
which is based on a result from \cite{ZRMT} that the expression
\[C_n\equiv\frac{i}{2}\sum_{m,j}\left(b_{n+m}^{j\da}b_m^j-b_m^{j\da}b_{m+n}^j\right)\]
commutes with any quasifree Hamiltonian of the form (\ref{Hqfree}). Writing down the expectation value in the ground state of 
a translation invariant Hamiltonian, the summation over $m$ is over 
identical terms and we have
\[\langle C_n\rangle =N \big\langle Im\sum_j b_{n+m}^{j\da}b_m^j\big\rangle\ \]
and since any translation invariant Bogoliubov transformation can be written as $U=e^{iH}$ with $H$ translation invariant
quasifree Hamiltonian, the invariance property (\ref{rinv}) follows.
 
This gives a rather simple way to check
its relation to the absence of the spectral gap, namely, we can look at the diagonal basis. 
The summed two-point function in the diagonal basis reads
\[\sum_j\langle b^{j\da}_m b^j_n\rangle  = 
\frac{1}{2N}\sum_{k,j}\exp\frac{2\pi ik(m-n)}{N}\left(-\frac{\Lambda_{-k}^j}{|\Lambda_{-k}^j|}+1\right)\] 
Its imaginary part is proportional to the k-antisymmetric part of the Fourier kernel:
\[\sum_j\frac{1}{2}\left(\frac{\Lambda_{k}^j}{|\Lambda_{k}^j|} -\frac{\Lambda_{-k}^j}{|\Lambda_{-k}^j|}\right)=\sum_j M_k^j\ ,\]
which vanishes unless $M^{j_0}_{k_0}\neq 0$ for a $j_0$ and $k_0$,  and this means the model is gapless as explained before.

\subsection{Consequences for mean-field approximations}

Our result does not hold for interactive systems, but it does for any translation invariant quasifree one, which
is possibly used in a mean-field approximation. Hence, in case we find that (\ref{rinv}) is non-vanishing for some $m,n$ 
in a gapped ground state of an interacting system, then the mean-field approximation based on the quasifree state with the
matching covariance matrix will not converge. More precisely, as a consequence of our result, the entanglement entropy of 
the latter state necessarily violates the area law and it has algebraically decaying correlations as opposed to the area 
law satisfying gapped ground state with exponentially decaying correlations it is supposed to approximate. The approximation obviously
cannot work. 

An example is the Majumdar-Ghosh model \cite{MG}, a nearest-neighbour and next-nearest-neighbour Heisenberg chain:
\[H=J\sum_j S_j\cdot S_{j+1}+\frac{J}{2}S_j\cdot S_{j+2}\] where 
$S_m \cdot S_n=\sigma^x_m\sigma^x_n+\sigma^y_m\sigma^y_n+\sigma^z_m\sigma^z_n$ in terms of Pauli matrices $\sigma_p$ at site $p$. 
Its two ground states are known to be the singlets $|\uparrow\downarrow\rangle-|\downarrow\uparrow\rangle$ between sites 
$2j,2j+1$ for all $j$ those between sites or $2j+1,2j+2$. The model can be rewritten in fermionic language via the Jordan-Wigner
transformation (see e.g., \cite{VHB}):
\begin{eqnarray}
H&=& J\sum_{l=1}^L \left[\bigg(\frac{1}{2}c_l^\dagger c_{l+1}
+\frac{1}{4}c_l^\dagger c_{l+2}
-\frac{1}{2} c_l^\dagger c_{l+1}^\dagger c_{l+1}c_{l+2}
\right.\nonumber\\
&+&h.c.\bigg)
+\left(c_l^\dagger c_l-\frac{1}{2}\right)
\left(c_{l+1}^\dagger c_{l+1}{-}\frac{1}{2}\right)
\nonumber\\
\bigg. &+&\frac{1}{2} \left(c_l^\dagger c_l-\frac{1}{2}\right)
\left(c_{l+2}^\dagger \,c_{l+2}{-}\frac{1}{2}\right)\bigg]
\label{jwham}
\end{eqnarray}
and rewriting the ground states also in the fermionic language, one 
finds that $\langle c_l^\dagger c_{l+1}\rangle=\langle c_{l+1}^\dagger c_l\rangle=1/4$.

Now, we can perform a transformation $c_l\mapsto c_l e^{i l \alpha}$ with a parameter $\alpha$, which multiplies 
the coefficients of the first three term in (\ref{jwham}) by $e^{i\alpha}$ and the second and third terms by $e^{2 i \alpha}$. 
More importantly the two-point function after the transformation is 
$\langle c_l^\dagger c_{l+1}\rangle=\langle c_{l+1}^\dagger c_l\rangle=1/4 e^{i\alpha}$. Now, 
the quasifree translation invariant model, which one would use to define the mean field (or generalised Hartree-Fock) 
approximation, with this property is gapless as we proved. Thus it has logarithmically diverging entropy and algebraically 
decaying correlations, it cannot approximate
the ground state of the gapped Majumdar-Ghosh model \footnote{This is why no complex solution for $\langle c_l^\dagger c_{l+1}\rangle$
was found by Verkholyak, Honecker and Brenig in \cite{VHB} for
the mean field equations.}.
\section{Acknowledgements} Big thanks to Giandomenico Palumbo for enlightening discussions. The work was supported by the EPSRC grant EP/I038683/1 and the
University of Leeds Acedemic Development Fellowship scheme.

\end{document}